\documentclass[twocolumn,showpacs,preprintnumbers,amsmath,amssymb]{revtex4}
\usepackage{graphicx}
\usepackage{dcolumn}
\usepackage{bm}
\begin{document}

\title{The Shortest Path Across the Mesoscopic System}

\author{Liqun He${}^{1,2}$, Eugene Kogan${}^{2}$, and Dawei
Luo${}^{3}$}

\affiliation{${}^1$Department of Thermal Science and Energy Engineering,\\
University of Science and Technology of China, Hefei, P.R.China}
\affiliation{${}^2$Minerva Center and Jack and Pearl Resnick
Institute of Advanced
Technology,\\
Department of Physics, Bar-Ilan University, Ramat-Gan 52900,
Israel}
\affiliation{${}^3$Department of Mechanical Engineering, \\
University of Kentucky, Lexington, KY 40506-0108, USA}

\date{\today }

\begin{abstract}
\leftskip 54.8pt \rightskip 54.8pt

We study distribution functions (DF) of mesoscopic hopping
conductance numerically by searching for the shortest path. We
have found that the distributions obtained by choosing randomly
the chemical potentials (for a fixed impurity configuration),
which corresponds to a typical experimental situation, coincide
with those obtained when both impurity configuration and chemical
potential is chosen randomly, in agreement with the ergodicity
hypothesis. The DFs obtained for one-dimensional systems were
found to be quite close to the independent predictions of V.I.
Mel'nikov, A.A.Abrikosov and P. Lee {\it et al}. For $D=2$, the
DFs both for narrow system and thin film look similar (and close
to the $1D$ case).The distribution function for the conductance of
the square sample is nearly Gaussian as predicted by both
Altshuler {\it et al} and Serota {\it et al}.

\end{abstract}


\maketitle
Mesoscopic conductance fluctuations in the insulating regime of
small, disordered transistors were first observed by
Pepper\cite{pep} in GaAs MESFETs and then studied in detail in Si
MOSFETs  by Fowler, Webb and coworkers\cite{fow} in the early
1980s. Extremely strong random fluctuations, spanning several
orders of magnitude, were observed at low temperatures in the
conductances of narrow-channel  devices as the gate voltage was
varied. The lognormal distribution for one-dimensional (1D)
conductors was long ago analytically predicted by V.I.
Mel'nikov\cite{Melnikov} and A.A.Abrikosov\cite{Abrikosov}. The
explanation was provided by Lee\cite{lee} who proposed a model in
which electrons move by variable-range hopping (VRH) along a 1D
chain. Serota, Kalia and Lee\cite{ser} went on to
simulate the ensemble distribution of the total chain resistance
$R$ and its dependence on the temperature  $T$ and the sample
length $L$. In their ensemble, the random impurities are
distributed uniformly in energy and position along the chain. In
experiments  a single device is generally used, so that the
impurity configuration  is fixed, and fluctuations are observed as
a function of some variable external parameter such as the
chemical potential. An ergodicity hypothesis is then invoked to
the effect that the same ensemble is sampled in both cases,
something that has been verified experimentally by Orlov {\it et
al.}\cite{orl1}. Using the natural logarithm of the resistance,
the authors of Ref. \onlinecite{ser} obtained for the mean and
standard deviation:

\begin{equation}
\label{1}
\langle \ln\rho \rangle \sim \left(\frac{T_0}{T}\right )^{1/2}
\left [\ln\left(\frac{2L}{\xi}\right)\right]^{1/2}
\end{equation}
\begin{equation}
\label{2}
s\equiv\langle \left(\ln\rho- \langle \ln\rho \rangle\right )^{2}
\rangle \sim \left(\frac{T_0}{T}\right )^{1/2} \left
[\ln\left(\frac{2L}{\xi}\right)\right]^{-1/2}
\end{equation}
where $\xi$ is the localization radius and $T_{0}$ is the
characteristic temperature for Mott VRH:  $T_{0}~=~1/k_{B}\rho
\xi$ ($\rho$ is the density of states at the Fermi energy). It can
be seen that the size $s$ of the fluctuations decreases extremely
slowly with  length, a result characteristic of 1D which was first
pointed  out by Kurkijarvi \cite{kur}. The explanation is simply
that exceptionally  large resistance elements, even though they
may be statistically  rare, dominate the overall resistance since
they cannot be by-passed  in this geometry. The averaging assumed
in the derivation of  Mott's hopping law for 1D does not occur and
the total resistance takes on the activated form of the largest
individual element.
 where $\xi$ is the localization radius and $T_{0}$ is the
characteristic temperature for Mott VRH:  $T_{0}~=~1/k_{B}\rho
\xi$ ($\rho$ is the density of states at the Fermi energy).

The crossover from 1D to 2D were also studied
\cite{Serota,Igor,XCXie}. The first theoretical description of
hopping conductivity in narrow 2D strips was given by
Serota\cite{Serota}, and numerical simulations from narrow 2D to
square 2D at a certain temperature were also done by X.C. Xie and
S. Das Sarma\cite{XCXie}. The conductance DF (on the metallic
side) was first fully considered analytically in 2D and above by
Altshuler, Kravtsov and I. V. Lerner\cite{Igor}, where they have
predicted the Gaussian distribution with long lognormal tails. Due
to the finite widths of channels, the 2D VRH is also affected by
the width of system. To observe this effect fully, L. He {\it et
al}\cite{vrh} did numerical simulations on all 2D cases, including
the short 2D which has been studied experimentally by Hughes {\it
et al}\cite{Hughes}. Although most of the results are in
accordance with these of the previous work, the DF for short 2D is
of a long tail to the low conductance opposed to the low
resistance as expected. To check the results, we numerically
simulate these samples by searching for the shortest path across
system other than by percolating it. The shortest paths would be
punctures which short out less conductive paths in the 2D
geometry. The aim of the present work is to report this study, and
first of all, it is also started by replacing the transport
problem across a 2D mesoscopic system in Mott hopping regime with
a random resistor network as did previously.

Imagine that a particle is about to transport a conductive
network. In the principle of lowest energy, it prefers to the path
of the smallest overall resistance among all possible paths. The
path of lowest energy cost is equivalent to the shortest path of
graph theory, which is just what the nature of puncture of the
system means. It means that we could also approximate the
resistance of the system by looking for the shortest path, as well
as by percolating the system as usual. The
shortest paths among percolation systems have been studied
considerably recently \cite{Nehemia,Dok,Gerald, Stanley1,
Stanley2} in terms of minimal path or optimal path between a pair
of sites within the same cluster, and the study on the scaling
form of the probability of the shortest path with regard to their
Euclidean distance and the cluster mass ($M_B$)\cite{Gerald} has
shown that the average conductance of the percolation backbone is
strongly correlated with the shortest path, and it decreases with
increasing minimal path. This means the shortest path determines
the average conductance in nature.

A network is set up by resistors, $\rho_{ij}$, between sites $i$
and $j$
\begin{equation}
\label{eqnRij} \ln\rho_{ij}=2\alpha d
+(|E_i-\mu|+|E_j-\mu|+|E_i-E_j|)/2kT ,
\end{equation}
Here 
$\alpha$ is the inverse localization length,  $d$ is the distance
of two localized sites, $E_i$ and $E_j$ are energies of site
$i,j$, and $\mu$ is the chemical potential, $T$ is the
temperature. Energy is chosen randomly from a uniform distribution
in the range $-0.5 \sim +0.5$. Thus the mesoscopic system is
reduced to a random resistor network(RRN). To percolate the
network, the resistors joining electrodes are selected in
ascending order until the first percolation path connects the
reservoirs. The resistance of the percolation path is taken to be
the resistance of the entire system.

To solve it by the shortest path, Dijkstra algorithm\cite{Thomas}
is applied, which is efficient in finding shortest paths to all
nodes from a single source in a fully connected graph. In views of
graph, the RRN is partially connected, and there are more than one
source nodes except for the 1D case. So, some modifications on
this algorithm has been done in the simulation.

In calculation, positions of impurities are uniformly distributed
over the system, their energies are distributed evenly between
$-0.5 \sim +0.5$, and gate voltage, $\mu$, is randomly chosen.
Thus we can consider the chemical potential distributions (for a
fixed impurity configuration) and the ensemble distributions( for
a fixed chemical potential ). For a 1D system of $L=1000$,
chemical potential $\mu=0$ and temperature $T=0.001$, we have got
the profile of their individual resistances similar to that along
percolating path(Fig.1 of Ref.\cite{vrh}). It means that it is
also the single largest hop along the shortest path that controls
the overall conductance of the system. With temperature
increasing, the sizes of fluctuation of individual resistors along
the shortest path become close.
\begin{figure}
\includegraphics[height=6cm,width=4cm]{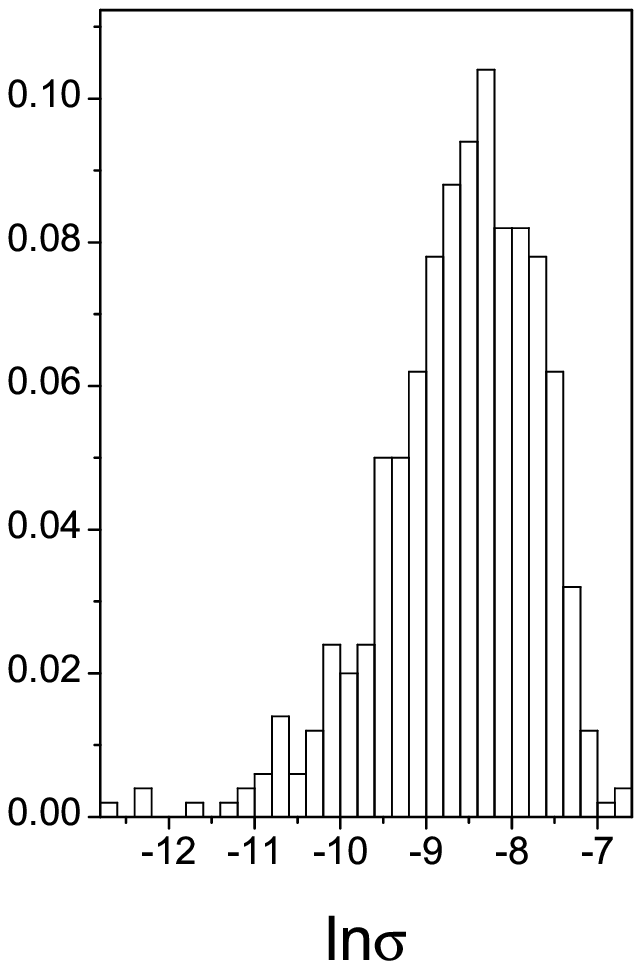}\includegraphics[height=6cm,width=4cm]
{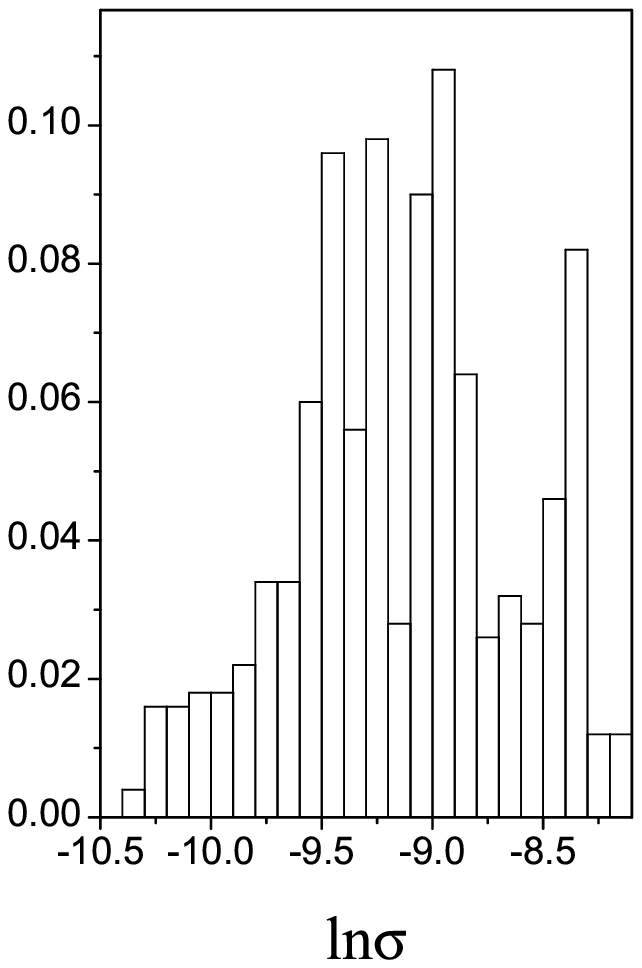} \caption {The conductance  of 1D system for
$\nu=0.225$: a) ensemble distribution function; b) chemical
potential distribution function. The histograms in both figures
are the numerical result}
\label{figdf}
\end{figure}
 As has been done previously\cite{vrh}, we also simulate the 1D
systems firstly to validate the simulation. The results are
shown in Fig.(\ref{figdf}). The size of 1D system is 1000 in
length, and 50 in localization length. The temperature is $T=
0.001$, which gives  $\nu=0.225$.
The chemical potential range is $\mu=-0.1\sim+0.1$. As found in
Ref.\cite{vrh}, with the increase of temperature, the fluctuation
of 1D resistance become small.

\begin{figure}
\includegraphics[height=6cm,width=4cm]{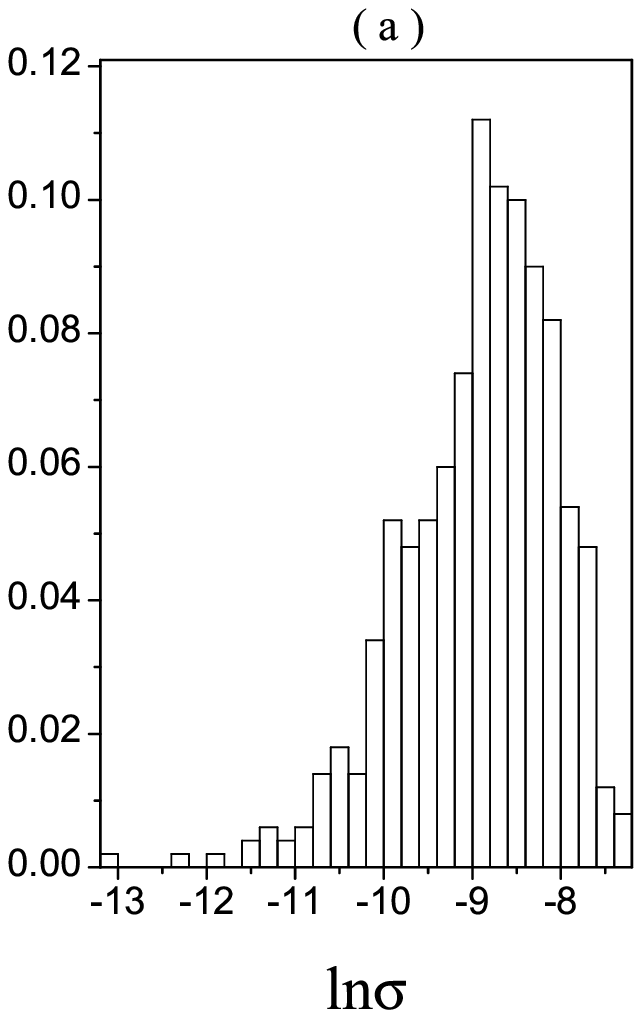}\includegraphics[height=6cm,width=4cm]{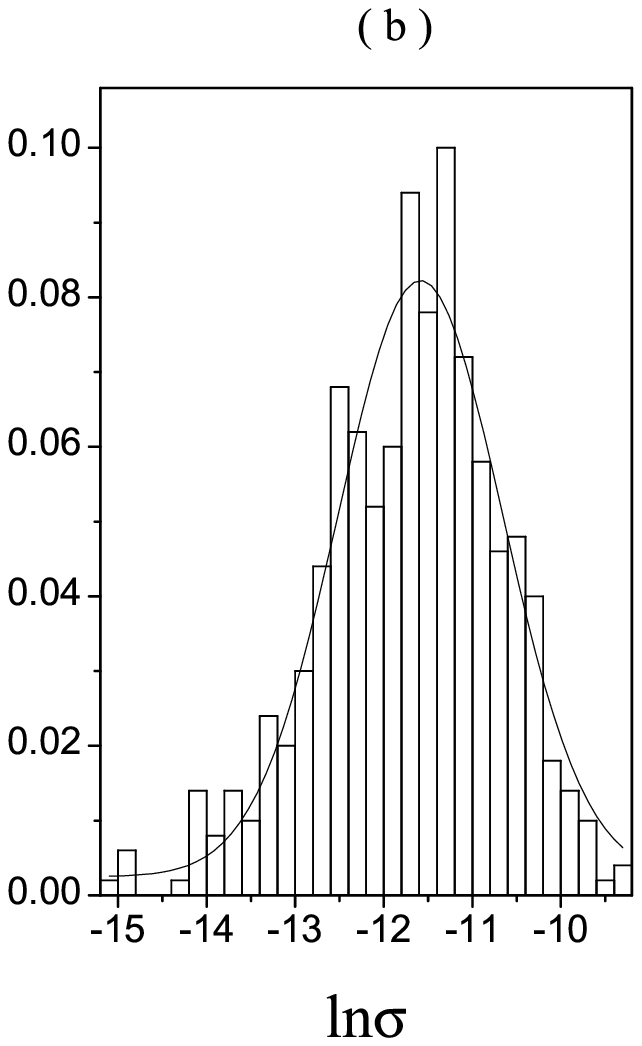}
\includegraphics[height=5.5cm,width=4cm]{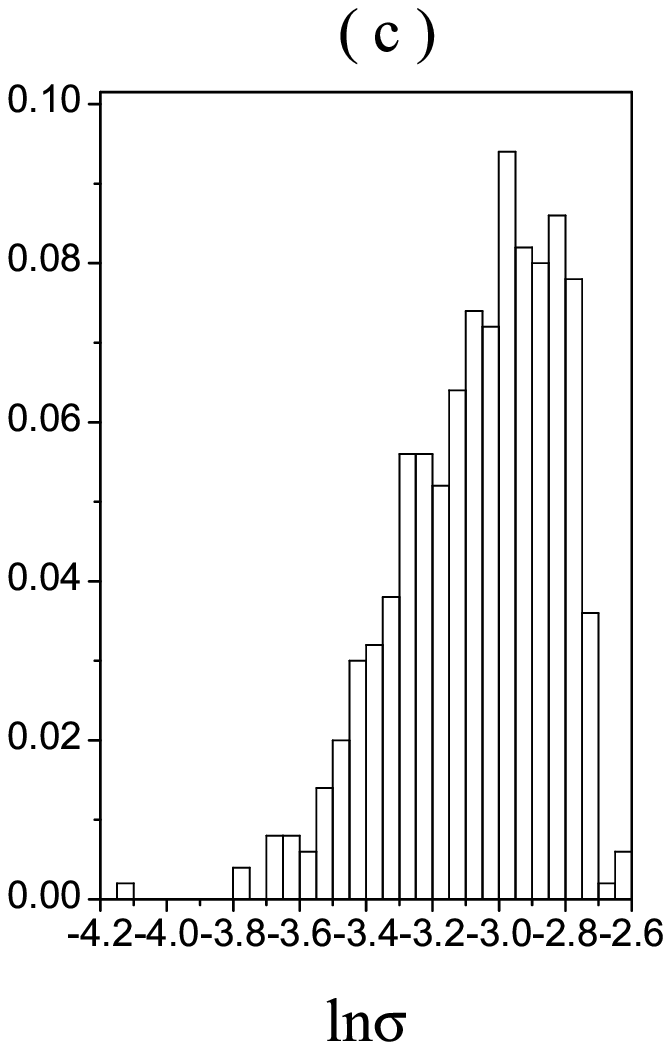}
\caption{The ensemble distribution function for the conductance of
2D system: a) long system; b) square sample; c) short 2D.
Histograms are numerical results. The solid curve in a) is the RR
theoretical fitting for the 1D case, and in b) is the Gaussian
fitting.} \label{DF2D}
\end{figure}

On the basis of the above validity, the 2D cases are numerically
simulated by the shortest path, including cases of narrow 2D,
square 2D and short 2D. In our simulation, localization length,
temperature,{\it etc.}, for a 2D system, $w \times L$, are the
same as in the 1D case, while $w=100, L=1000$ for narrow 2D,
$w=1000$ and $L=100$ for short 2D  and $w=L=1000$ for square 2D
system. The results are shown in Fig.\ref{DF2D}. From this, it is
seen clearly that the magnitude of conductance substantially
increases with the increase of system width, which is in agreement
with the prediction of both Serota \cite{Serota} and Altshuler
{\it et al} \cite{Igor}, and the numerical results of X.C. Xie
{\it et al}\cite{XCXie} for narrow 2D and square 2D cases. The
results also reveal that the situations of narrow 2D is close to
the 1D case, and the normal 2D is close to Gaussian. As discovered
previously\cite{vrh}, the DF of conductance across the short 2D
system is still close to that of 1D case other than a mirror
reflection of the latter as expected. Note that the peak of short
2D is shifted to the low end of resistance more than the other 2D
cases. According to this, it seems to be more proper to explain
the puncture nature of 2D systems by the position of the peak of
DF while the long tail of DF shows the bias conductance
fluctuations.

In conclusion, the paper has studied the DF of  the conductance in
mesoscopic systems by the shortest path which could reflect the
puncture nature of VRH system more directly. The validity is done
and the results are found to be quite close to the numerical
results of the narrow and square $2D$ systems by P.A.
Lee\cite{lee}, Serota\cite{ser} and X.C. Xie {\it et
al}\cite{XCXie}, \emph{i.e.}, the result for narrow $2D$ system is
close to that of $1D$ case, and the result for square $2D$ is
close to Gaussian\cite{Igor}. But the unexpected results for short
2D system hints that the puncture nature of 2D systems could be
explained more properly by its peak position of DFs and the long
tail shows the preference of conductance fluctuation.

\begin{acknowledgements}
Authors acknowledge valuable criticism of I. V. Lerner. One of the
authors (He) thanks Dr. N. Schwartz for his introduction to the
Dijkstra algorithm, and also thanks Prof. Shlomo Havlin for his
hospitality in BarIlan University of Israel. This research was
supported by THE ISRAEL SCIENCE FOUNDATION founded by The Israel
Academy of Sciences and Humanities, NSF of Anhui Province, China
(00047520) and THE STARTING FOUNDATION FOR RETURN SCHOLARS by
Chinese Government(2001-2002).
\end{acknowledgements}


\begin{thebibliography}{99}

\bibitem{pep} M. Pepper, J. Phys. C {\bf 10}, L173 (1977).
\bibitem{fow} A.B. Fowler, A. Hartstein, and R.A. Webb,
 Phys. Rev. Lett. {\bf 48}, 196 (1982);
 R.A. Webb, A. B. Fowler, A. Hartstein,
and  J. J. Wainer,  Surface Science {\bf 170}, 14 (1986). V.I.
\bibitem{Melnikov}Mel'nikov, Sov.Phys.Solid State (FTT), v.23,
441 (1981)
\bibitem{Abrikosov}A.A.Abrikosov, Sol.State Commun., v.37,
997 (1981).
\bibitem{lee} P. A. Lee,  Phys. Rev. Lett. {\bf 53}, 2042 (1984).
\bibitem{ser} R. A. Serota, R. K. Kalia, and P. A. Lee,
 Phys. Rev. B {\bf 33}, 8441 (1986).
\bibitem{orl1} A. O. Orlov, A. K. Savchenko, and A. V. Koslov,
Solid State Commun. {\bf 72}, 743 (1989).
\bibitem{kur} J. Kurkijarvi, Phys. Rev. B {\bf 8}, 922 (1973).
\bibitem{Serota}
R.A. Serota, Solid State Communiations, Vol {\bf B 67}, No.11,
pp.1031-1033,1988.
\bibitem{Igor} B. L. Altshuler, V. E. Kravtsov, and I. V.
Lerner, JETP Letters, 43, 441;  ZhETF 91, 2276
\bibitem{XCXie}
X.C. Xie and S. Das Sarma, Phys. Rev {\bf B 36}, 4566 (1987).
\bibitem{vrh}Liqun He, Eugene Kogan, Moshe Kaveh, Shlomo Havlin, and Nehemia Swartz, cond-mat/9911299.
\bibitem{Hughes}
R.J.F.Hughes, A.K.Savchenko, J.E.F.Frost, E.H.Linfield,
J.T.Nicholls, M.Pepper, E.Kogan and M.Kaveh, Phys. Rev {\bf B 54},
2091 (1996).
\bibitem{Nehemia}Nehemia Schwartz, Markus Porto, Sholomo Havlin, Armin Bunde, Physica A 266(1999) 317-321
\bibitem{Dok}Nikolay V. Dokholyan, Youngki Lee, Sergy V. Buldyrev, Sholomo Havlin,  Peter R. King, and H. Eugene Stanley,PHYSICA A
 266 (1-4): 55-61 APR 15 1999
\bibitem{Gerald}Gerald Paul,Sergy V. Buldyrev, Nikolay V. Dokholyan, Sholomo Havlin, Peter R. King, Youngki Lee, and H. Eugene
Stanley, cond-mat/9910236v1, 15 Oct 1999
\bibitem{Stanley1}H. Eugene Stanley, Jose S. Andrate Jr., Sholomo Havlin,Hernan A. Makese, Bela Suki, Phusica A 266(1999) 5-16
\bibitem{Stanley2} H. Eugene Stanley, Jose S. Andrate Jr., Physica A 295(2001) 17-30
\bibitem{Thomas}Thomas H. Cormen, Charles E. Leiserson, Ronald L. Rivest,{\em Introduction to Algorithm},The MIT Press, 1994.
\end{thebibliography}
\end{document}